# Conceptual Modeling of the Whole–Part Relationship

Sabah Al-Fedaghi[*]

*Computer Engineering Department*
*Kuwait University*
*Kuwait*

salfedaghi@yahoo.com, sabah.alfedaghi@ku.edu.kw

*Abstract -* **Conceptual models rely on structural information to describe relationships among UML classes; among these, the *whole–part (WP) relationship* plays a fundamental role. This paper explores and analyzes the WP semantics at large with a focus on its software engineering use. The WP relationship has often been treated as a first-class modeling construct in object-oriented analysis, a subject of keen interest and it is considered important for UML modeling. From the scientific and philosophical aspects, a theory of parts forming a whole is a complex issue, loaded with controversies that are widely discussed. This paper aims to offer a semantic *assembly* model that is useful to describe WP relationships in conceptual modeling. We contribute to the WP research by conducting an ontological analysis using UML samples that exemplify the WP construct. The method of investigation is based on a model called a thinging machine (TM) to explore the WP semantics through applying TM to numerous existing UML models. The TM model uses the so-called thimacs (things/machines) to form building blocks for describing the domain at a three levels of description: static, events, and behavioral models. This approach contrasts the UML method, which is infected by a multiplicity problem concerning the integrated view of structure and behavior and how to associate diagrams with one another. This investigation's results point to a promising contribution to the understanding of the notion of WP relationship in UML.**

## I. INTRODUCTION

Conceptual models are artifacts produced with the deliberate intention of describing a conceptualized reality that incorporates an idealistic view of an existing domain in the world with an objective account [1]. A conceptual model always has a conceptual semantics so human conceptualization and perception mediates the reference to the world [2]. Conceptual modeling pertains to identifying, analyzing and describing the essential concepts and constraints of a domain with the help of (diagrammatic) modeling language [3]. It helps in understanding and communicating among the stakeholders and serves as a base for consequent phases of a system's development [4].

According to Catossi et al. [5], the difficulty of building liable conceptual models occurs due to the lack of domain knowledge. There are some ontological analysis techniques that can help in such a task. However, they are not easy to use because they involve many philosophical concepts, which makes them complex to the common modeler. A conceptual model that does not represent the domain adequately can lead to problems with respect to systems project, implementation, operation and maintenance [5]. UML on its own cannot guarantee a model that is free of conceptual mistakes. Attempts to improve these situation techniques based on ontological analysis have been explored to help modelers validate their UML class diagrams more easily [5].

Structural information emphasizes the system's static structure using objects, attributes and relationships. Structural information usually has the semantics of classification, generalization and aggregation to describe structural relationships among classes; among these relationships, a fundamental role is played by the so-called *whole–part* relation. In this paper, structure is applied to an ontological unit that has a dual composition of a thing and a machine. Hence, structure refers to dual structural part/whole aspects, one for a thing and one for a machine.

### A. About the Whole–part Relationship

This paper focuses on the semantics of the whole–part notion and extends analyses of the semantics of UML's aggregation and composition to behavior. We address the development of semantics to the whole–part notion by specifying a diagrammatical-based representation called thinging machines (TM). In UML, the 'whole part' (the whole may be called *assembly*) relationship is viewed regarding aggregations, which are special associations. In TM, the 'whole part' is expressed as a relationship between the so-called thimac (*thi*ng/*mac*hine) and subthimacs.

The concept of whole–part structures has been a research issue for ontology, cognitive science and linguistics. Such a theory (under the name: mereology) concerns the investigation of the ontology of whole–part relationships, or relationships of part to whole and the relations of the part with another part within a whole [5]. According to Guizzardi [6], parthood is a relation of fundamental importance in conceptual modeling because it is present as a modeling primitive in practically all major conceptual modeling languages. Motivated by this, numerous attempts have been made to employ theories of different sorts to provide a foundation for whole-part relationships.

* Retired June 2021, seconded fall semester 2021/2022

The whole–part relationship has often been treated as a first-class modeling construct in object-oriented (OO) analysis, design methods and OO modeling languages [7]. According to Barbier et al. [8], the whole–part relationship in OO modeling has been a subject of keen interest and is considered important for OO modeling and in UML.

### B. Problem

A theory of parts forming a whole is a complex issue and loaded with many controversies; its scientific and philosophical aspects are widely discussed. There is much controversy over some properties involved in the whole–part relationships because they result in being inconsistent with the conceptual theories applied in the real world. Problems occur due to ambiguities about the concept of parts in addition to cases in which objects of a completely different nature could be considered as whole–part relationships [5]. According to Barbier et a. [7], "This concept remains somewhat vague in OO due to, in particular, the wide range of terms used in the literature (e.g., Aggregation, Composition, Assembly, Containment, Membership)." The way the whole–part relationship is formalized is unsatisfactory [8]. In OO systems, the representation of whole–part relationship usually requires a particular semantics, but rarely do current modeling formalisms and methodologies give it a specific "first-class" dignity [9].

In this article, we research the whole–part relationship by conducting an ontological analysis to investigate its representations in UML. Currently, there has been a rising interest in applying ontologies to develop real-world semantics for conceptual modeling and methodological guidelines for improving conceptual modeling.

### C. About This Paper

In Section 2, we briefly review the basics of thinging machines and include some new discussion about TM modeling. There are many papers about such a topic; we refer to one old and one recent samples [10-11]. In Section 3, we introduce an example that illustrates the TM modeling and its application in UML The remaining sections of the paper address several issues including the association relationship in UML versus TM, object thimacs, the whole–part relationship, priority of the whole versus part and material objects.

### II. THINGING MACHINE MODEL

The TM modeling is based on a category called thimacs (things/machines), which is denoted by Δ. The Δ has a dual mode of being: the machine side, denoted as M, and the thing side, denoted by T. Thus, Δ = (M. T.) – see Fig. 1. In TM modeling, thingness and machinery cannot be separated, but it is often convenient to focus on one or the other aspect. A thimac is similar to a double-sided coin (see Fig. 2). One side of the coin exhibits the characteristics the thing assumes, whereas on the other side, potential actions emerge as a machine. A thing is subjected to *doing*, and a machine *does*, as shown in Fig. 3.

The *thing* embodies a fixed background "space" while the *machine* represents the roots of possible dynamism. The TM model is constituted by static thimacs, as well as by higher-level thimacs when time is injected in the static thimacs.

The TM machine has five actions: create, process, release, transfer and receive (Fig. 4). The structure of the thimac as a thing and a machine is further illustrated in Fig. 5.

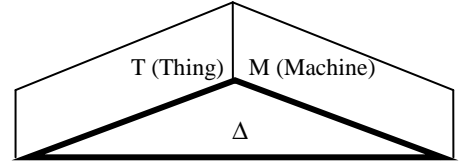

Fig. 1 A thimac has a dual mode of being a thing and a machine

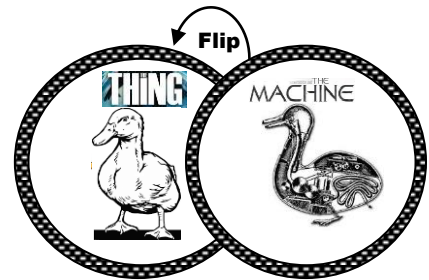

Fig. 2 Illustration of the duality of a thimac

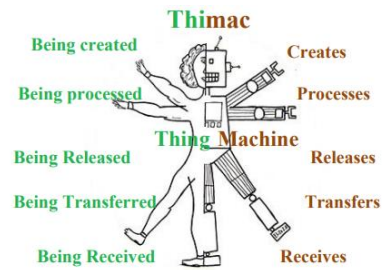

Fig. 3 Thimac

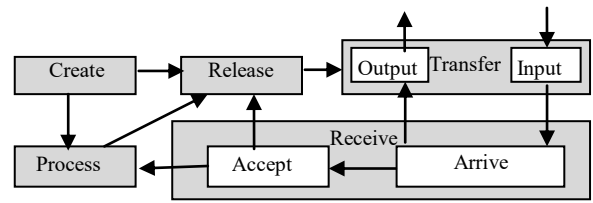

Fig. 4. Thinging machine

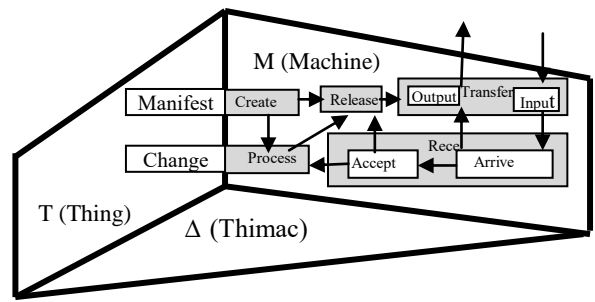

Fig. 5. Illustration of the thimac as a thing and a machine.

## A. Example of Thimacs

To clarify the notion of a thimac, we construct the thimac that corresponds to tuples. According to Waguespack [12], the most concrete concept in the relational paradigm is the tuple, which corresponds one-one with *a single concept of reality*. Fig. 6 shows a tuple as a thimac. In this tuple (circle 1), the attribute values-things (2 – we use UML terminology for illustrative purposes) flow to the tuple machine (3), where they are received (4) and processed (5). Such a processing (e.g., concatenation of values) triggers creation (6), which is manifested (7) as a whole tuple thing (8) that represents, in Waguespack's [12] words, "a single concept of reality." Now, the tuple can be released, transferred, received and processed (tuple's machine) as a thing.

The assemblages of thimacs are formed from a juxtaposition of subthimacs that are bonded into a structure at a higher level at which they become parts. Thimacs comprise parts, which themselves are thimacs that comprise parts, and so on (see Fig. 7). Thimacs cannot be reduced to their parts because they (wholes) have their own machines.

## B. The Machine

The term *machine* reflects the thimac's universal implication that *everything is a machine* (as a thimac), even a poem. The American poet William Carlos Williams wrote in 1944, "A poem is a small (or large) machine made of words"; and "When we say 'This is where the poem really starts to move,' 'Your poem picks up steam after these lines,' or 'What is this poem doing?' we're using the 'poem as machine' metaphor" (see a sample TM modeling of poems in [13]). The universal sense of machinery originated in the TM *actions* indicating that everything that creates, changes (processes) and moves (release-transfer-receive) is a machine. TM actions in Fig. 4 can be described as follows:

**Arrive**: A thing moves to a machine.

**Accept**: A thing enters the machine. For simplification, we assume that arriving things are accepted; therefore, we can combine **arrive** and **accept** stages into the **receive** stage.

**Release**: A thing is ready for transfer outside the machine.

**Process**: A thing is changed, handled and examined, but no new thing results.

**Create**: A new thing "comes into being" (is found/manifested) in the machine and is realized from the moment it arises (emergence) in a thimac. Such a description of "being" echoes Quine's slogan, "To be is to be the value of a variable." Existence in such a model refers to an internal existence, which is based on the relevant framework's rules. This thought is grounded in Carnap's philosophy that, "points to the existence of multiple types of logical frameworks we might construct, investigate, and apply for different purposes […] in analogy with the existence of different types of geometry" [14]. Note that for simplicity's sake, we omit *create* in some diagrams, assuming the box representing the thimac implies its existence (in the TM model).

**Transfer**: A thing is input into or output from a machine.

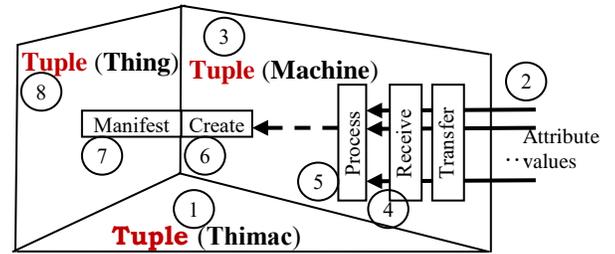

Fig. 6. The thimac tuple has a dual mode of being a thing and a machine.

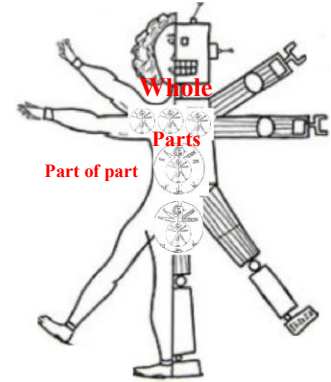

Fig. 7 A thimac and subthimacs

Additionally, the TM model includes the *triggering* mechanism (denoted by a dashed arrow in this article's figures), which initiates a (non-sequential) flow from one machine to another. Multiple machines can interact with each other through the movement of things or through triggering. Triggering is a transformation from the movement of one thing to the movement of a different thing. The TM space is a structure of thimacs that forms regions of events (to be defined later).

## C. Thimacs and Objects

The TM model is the whole system of thimacs, which is a grand thimac. The basic structure that represents the thimac/subthimacs in TM now reflects the whole–part relationship instead of nested objects (objects that are inside another object). Accordingly, we have to examine the subthimacs' roles as parts of the whole thimac. We can identify two extreme models. As shown in Fig. 8, the subthimac can have independent interactions with other thimacs (the whole, other subthimacs and outside thimacs). The other extreme case is when the whole thimac controls the interactions of its subthimacs. This thimac will be called an object thimac (othimac). Many heterogeneous thimacs exist between these two extreme cases. In general, othimacs can have *object subthimacs*.

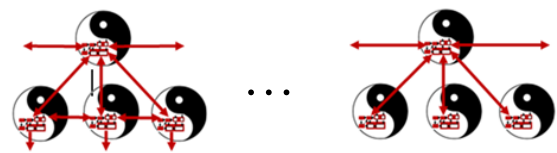

Fig. 8 Subthimacs interact with each other and the outside (left) and subthimacs interact only through their thimac (right)

## D. UML vs. TM modeling

Because most of our treatment of the whole–part modeling involves examples from UML, we discuss some modeling aspects in TM and UML. In UML, different models describe the static level (class diagram), instance level (object/instance diagram) and behavioral level (e.g., state diagram). Each level utilizes different types of notation. In UML, one approach to the design for a particular problem precedes as follows.

You would start by determining a complete list of classes, then determine all the associations, then fill in all the attributes, and so on. To discover whether an association should exist, ask yourself if one class *possesses control, is connected to, is related to, is a part of, has as parts, is a member of or has as members* some other class in your model." [15]

In TM modeling, the design for a particular problem starts with "picking up" a thing (thimac) and follow its flow through other thimacs until it triggers another thing that leads to a new stream of flow, and so on. For example, the flow of an order leads to the flow of an invoice, which leads to a flow of payment, which leads to the flow of product inventory. The complete modeling process involves three levels of description: static, events and behavioral models.

A static description provides complete descriptions of machines and inter-machines in the modeled system. The description constructs the whole model as a machine. To inject dynamics, we divide the static into regions. Identifying these regions produces the dynamic model. To realize changes, we inject time into these regions to create the events model (see Fig. 9). Finally, the events gather into a chronological order to form the behavioral model.

## III. TM MODELING: CASE STUDY

Catossi et al. [5] give a case study that includes the classes *customer*, *order* and *item* (numbers 1–3, respectively) as shown in Fig. 11. The *customer* class is viewed as the whole and *order* class as the part. One must identify if there is a generic dependence between those classes, that is, if the whole must have a part that can be substituted over time. According to Catossi et al. [5], the answer is "no" because if it is taken into account that the client register must stay stored, then the customer continues being a client without doing an order.

### A. Static Model

Fig. 11 shows the corresponding static TM model. Here, we introduce sample functions that include the modules A to F. Modules A, B and D are developed just to familiarize with the TM model. Other modules C, E and F in the figure demonstrate some whole–part links between thimacs (classes). The functional and static descriptions of these modules are as follows.

**A**: Constructing a record of a new customer and adding it to the customers file.

**B**: Retrieving a customer record.

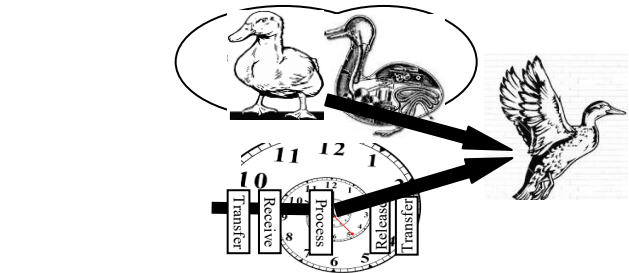

Fig. 9 Static and time thimacs generate dynamics.

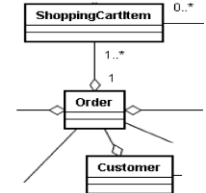

Fig. 10 Example from Catossi et al.'s [5] (partial)

**C**: Retrieving orders from a certain customer.

**D**: Retrieving a certain order.

**E**: Deleting a certain shopping cart item and its order.

**F**: Adding an order to a certain item.

Note that C links *customer* and *order* classes and E and F link *item* and *order* classes.

**Module A:** As shown in Fig. 11, this module involves receiving the values of attributes of a *customer* (4), constructing the *customer's record* (5) and then inserting the *record* in the *customers file* (6). Note that for the sake of brevity, we do not detail how the record is inserted in the file and describe the procedure as a mere process of the record and file.

**Module B**: This module involves outputting a certain *customer's record* when given a *customer ID* (7), which is assumed to be one the record's attributes. Again, we represent extracting the *record* from the *customer* as processing (8) the customer number and the file (9). It is possible to describe such a process in a further elaborate manner (e.g., sequential search loop over all records in the file).

**Module C**: This module demonstrates the link between the two classes (to use UML terminology), *customer* and *order*, and details extracting data from a file. The function here is retrieving a *customer record* from the *customers file* then finding all orders belonging to that *customer* (e.g., for daily report of customers vs. their orders). We will not show the loop for all customers.

Accordingly, processing the customers file extracts the *customer record* (10). Note that this extraction is modeled in terms of *transfer-receive* because the *customer record* is already created inside the **customers file**. Additionally, the *customer ID* is extracted from the customer record (11).

On the other side, the *orders file* (12) is processed (13) to extract an *order record* then a customer ID (14). This customer ID is compared (15) with the *customer ID* previously extracted from a *customer record* (11). If the two *order IDs* are the same, the order record is added to the *customer orders file* or the file of *order of a specific customer* (16).

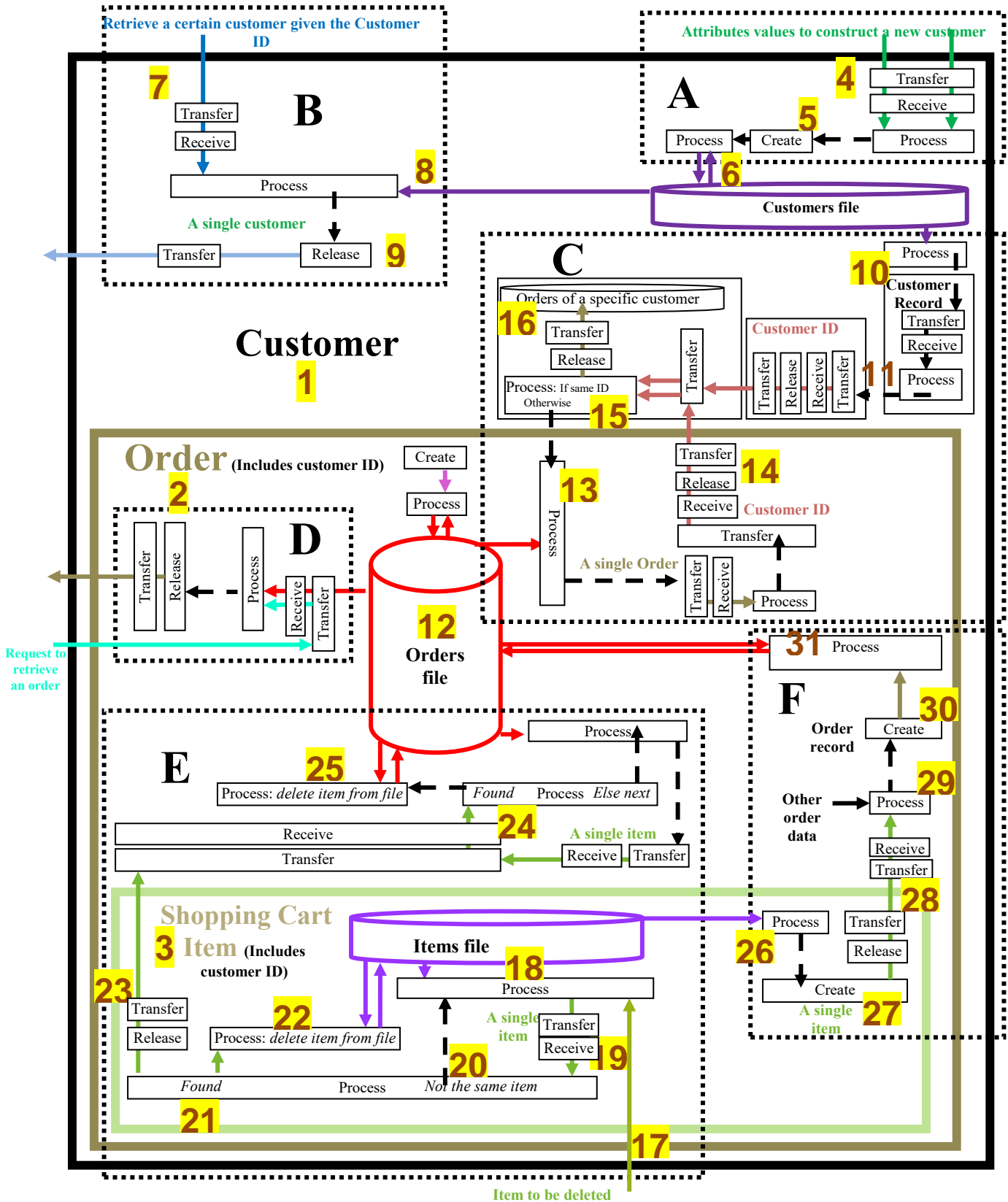

Fig. 11 The static model that corresponds to Catossi et al.'s [5] example.

Here, we assume a sequential search processing of *all* orders in the *orders file of customers*. The loop of this process will be specified in the events model.

**Module D:** This module is similar to Module **A** applied to retrieving an order; hence, it will not be discussed.

**Module E:** This module involves deleting an *item*, which causes deleting an *order* but deleting an *order* will not affect a *customer*. First, a request to delete an *item* is received (17). Just as in Module C, we assume a sequential search for the deleted *item* in the *items file*. Thus, the *items file* is processed (18) to extract an *item* (19). For the sake of simplification, we do not differentiate here between the item record and the item ID as we did in C.

Accordingly, if the item is not the item to be deleted (20), another item is retrieved from the file. Otherwise, if it is the item to be deleted (21), then:

- Delete the item from the items file (22). This deletion is modeled as downloading the file then uploading it again. It is possible to elaborate such a process by creating a new file that includes all read records except the deleted one.

- Send the deleted item record (which is the item ID) to the *order* class to delete the corresponding order (23).

In the order class, there is another sequential search loop for the corresponding order record (24). The order record is deleted when it is found (25).

**Module F:** This module demonstrates the link between the item class and the order class, assuming there is a one–one relationship between them. Hence, from the items file (26), an item record is retrieved (27) and sent to order (28). The item and the order data are processed (29) to create an order record (30) that is added to the ordered file (31).

### B. Events Model

To model the behavior of the static model in Fig. 11, we need to identify different events. A TM event is represented by a subdiagram of Fig. 11 (the event's region) and a time subthimac. For example, Fig. 12 shows the event *Inputting attribute values to construct a new customer* (in Module A). For the sake of simplification, the event will be represented by its region. To save space, we will a develop events model for only two modules C and E.

### C. Deleting an Item Events (Module E)

As shown in the events model Fig. 13, deleting an item involves the following events, where *vi* denotes event i.

v1: A request arrives to the item class to delete a certain item given an item ID.
v2: The items file is processed to retrieve the item record.
v3: The item record is not the required one.
v4: The item record is the required one.
v5: The item record is deleted from the file.

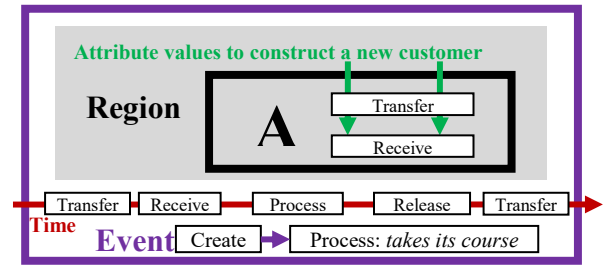

Fig. 12 The event *Inputting attribute values to construct a new customer*

v6: The item record (which includes the items order ID) is sent to the order class.
v7: The orders file is processed to retrieve an order record.
v8: An order record is retrieved from the file.
v9: The retrieved record is not the record to be deleted.
V10: The retrieved record is the record to be deleted; hence, it is deleted from the file.

Fig. 14 shows the behavioral model of *Deleting an item events (Module E)* regarding the time order of events. Note that we assume a sequential search of the *orders file*.

### D. Finding all orders of a given customer

As shown in the events model Fig. 13, retrieving all orders of a given customer involves the following events.

v11: A customer record is retrieved from the customers file.

v12: The customer ID is extract from the record.

v13: The *orders file* is processed to retrieve an order record.

v14: The customer ID is extracted from the order record.

v15: The two customer IDs are compared.

v16: The two customer IDs are not the same.

v17: The two customer IDs are the same; hence, the order record is added to the file of the customer's orders.

Fig. 15 shows the behavioral model for *Finding all orders of a given customer* regarding the time order of events.

Fig. 16 shows the basic TM relationships discussed in this example using UML-style notation. In such a figure, we abstract the TM diagrams 11 (static) and 13 (behavior).

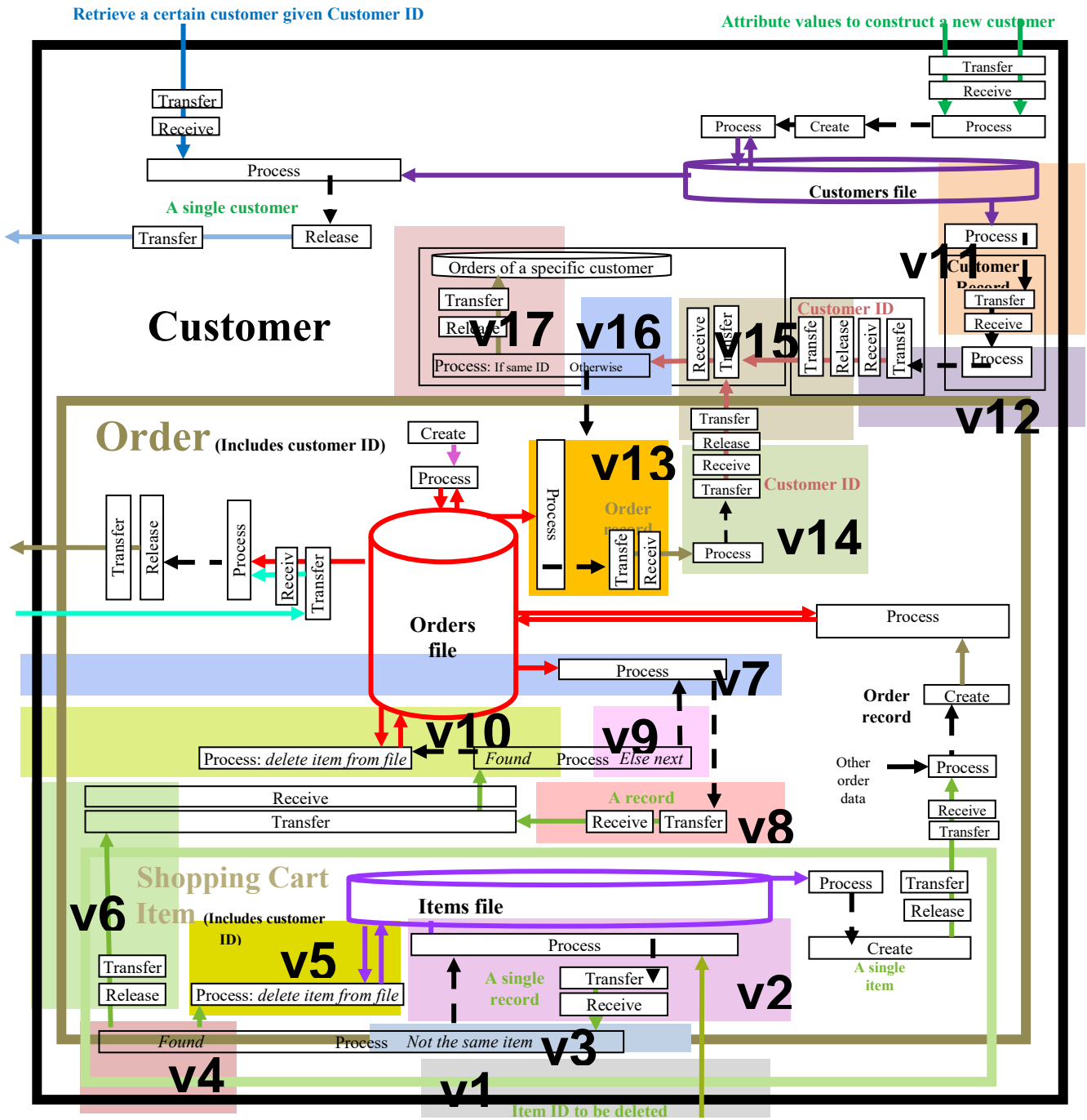

Fig. 13 The events model

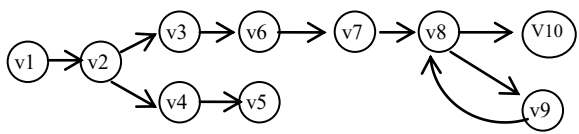

Fig. 14 The behavioural model of deleting an item and its order

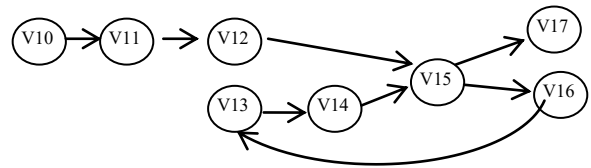

Fig. 15 The behavioural model of retrieving a certain customer's order.

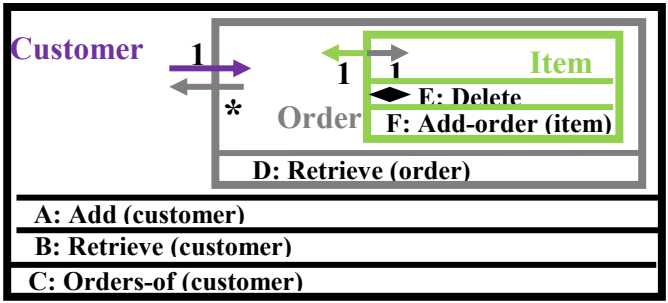

Fig. 16 UML-style basic relationships

## IV. ASSOCIATION: UML VS. TM

In UML, relationships clarify the way that elements interact or depend upon each other. For example, associations link model elements and indicate the nature and rules that govern the relationship. The basic way to represent association is with a line between the elements. Aggregation is a type of association relationship that specifies a collection of other elements form an element. A composition association relationship represents a whole–part relationship and is a form of aggregation.

Fig. 17 provides a UML example of association. It indicates that there must be exactly one *instance* linked to each object (instance) at the other end of the association, that is, there can only be one *company* associated with each *employee* [15]. Such a UML representation also involves *object* and *behavior* diagrams and may include other diagrams. The example (Fig. 17) demonstrates merely the static description because the behavior specification uses completely different notation. This approach raises a multiplicity problem concerning the integrated view of structure and behavior and how UML diagrams are associated with one another.

In TM, the different levels of modeling are built upon each other. Fig. 18 shows the TM static model that includes employee and company and the movement of employee to the company. The flow (red arrow) of employee to company represents the relationship. A relationship in TM, as everything else, is a thimac. Fig. 19 shows the TM events diagram that includes the instance employee, the instance company and the flow event from employee to company.

These events can be described as follows:
- There is a present (existing) employee $E_i$
- There is a present (existing) company $C_j$
- $E_i$ has been assigned to work in $C_j$

We assume that there are *multiple* employees and *n* companies. Additionally, this event of $E_i$ works for $C_j$ has the attribute (to use a UML term) *Uniqueness* that indicates the event happens once. A TM event is also a thimac, but it embeds a time subthimac. Thus, when an employee $E_i$ is assigned to a company $C_j$, he/she would not be assigned to any other company. Fig. 20 shows the behavior of this employee–company situation. Fig. 21 shows a possible tabular implementation. As an example, in this representation, the event uniqueness constraint is realized by searching in the table that $E_1$ has not already been assigned to a company.

As can be seen in this example, joining together different regulated (explicitly controlled) flows expresses the UML relationship notion in TM. There is no need for adding the relational concept in the model. We speculate that the process of establishing a relation goes through the three levels of static, events and behavioral subprocesses that is illustrated in Fig. 22.

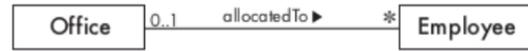

Fig. 17 Sample UML association (From [15])

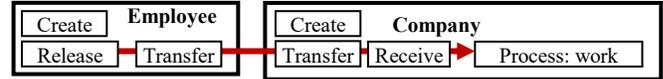

Fig. 18 The static model

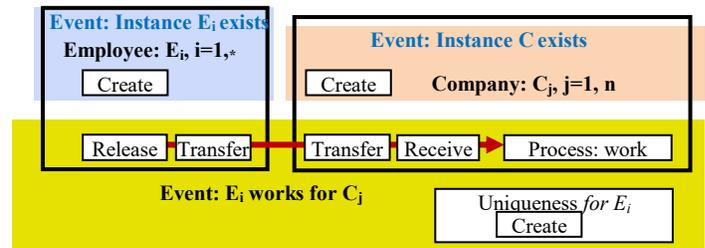

Fig. 19 The events (instances) model

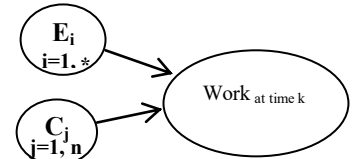

Fig. 20 The behavioral model

| Employee instance | Company entity | Event occurrence | |
|---|---|---|---|
| $E_1$ | $C_1$ | Work$_1$ | |
| $E_2$ | $C_1$ | Work$_2$ | |
| $E_3$ | $C_1$ | Work$_3$ | |
| $E_4$ | $C_2$ | Work$_4$ | |
| $E_1$ | $C_2$ | Work$_5$ | Not executed because $E_1$ already works in $C_1$ |

Fig. 21 Sample database

1. **Static recognition of thimacs and subthimacs** 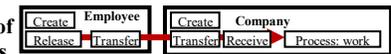

2. **Recognition of different events** 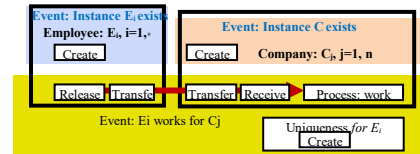

3. **Identifying the sequence of events** 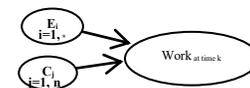

Fig. 22 The three levels of static, events and behavioral

## V. WHOLE–PART RELATIONSHIP AND OBJECT THIMACS

In this section, we relate the whole–part notion to the important distinction in the TM model between *object thimacs* (othimacs) and *non-object thimacs* (nthimacs). Simply, an othimac controls the handling (actions) of its parts when interacting with the outside of the thimac. Note that all thimacs (and subthimacs) have the same structure (TM machines). The structure is responsible for their behavior. The thimacs are similar to organisms in this uniform structural composition rather than being classical machines with structures that vary from one type to the other.

**Example**: Fig. 23 shows the static description of an *othimac* table. The static model represents the *type* table as in the UML class; however, it has richer semantics than a class does. The static model is an ontological assembly representation of all aspects in the life of the table (its construction, attributes, movement, etc.), except for time dynamics information. It is a union of all regions of events (construction, destruction, movement, modifications, features, new features, etc.). Then, such a picture of all regions of events is abstracted to delete all aspects not of interest in the current model frameworks.

In the current example of the table, we are interested in the composition of the table and the dimensions of its parts, color, its construction, and so on. However, we are not interested in where the nails used in its construction are made and other similar information. This static description may include contradictory information, such as its original color, which now has been changed. If we want to extract the process of constructing the table, then the events model (Fig. 24) produces the behavior of such a process, as displayed in Fig. 25. In Fig. 25, first the top surface and four legs are made, $E_1$, $E_2$–$E_5$, respectively. Then, the legs are nailed to the top surface ($E_6$–$E_9$) to form the table ($E_{10}$).

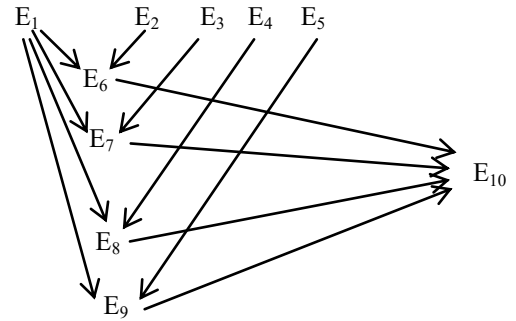

Fig. 25 The behavior model of constructing the table

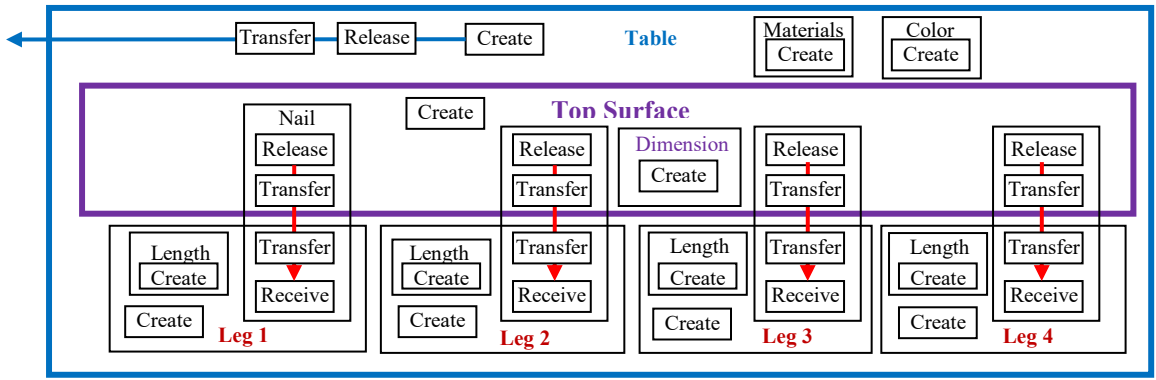

Fig. 23 Static model of a table

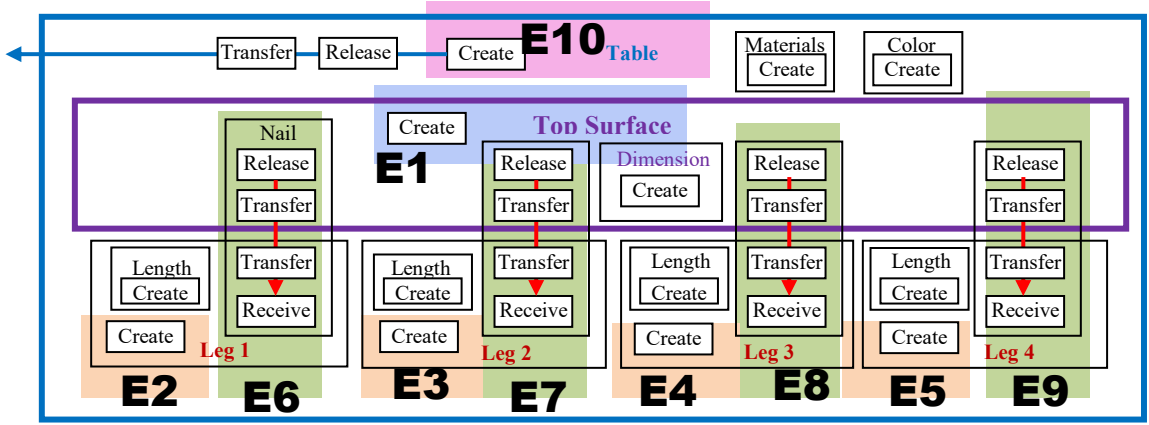

Fig. 24 The events model of constructing the table.

Fig. 26 illustrates the othimac whole idea using Matryoshka dolls' movement, which is performed only through the whole (the containing doll). Note that in these dolls, the part is *detached* from the whole; however, its movement is tangled with the actions of the whole. To illustrate this phenomenon in TM modeling, consider the following examples from Varzi [16].

1.  *The handle is part of the mug*. The mug/handle is an *othimac* because the handle movement (release/transfer) is tangled with the movement of the mug (see the TM model in Fig. 27). In this case, the handle is an *undetached component* of the mug.

2.  *The remote control is part of the stereo system*. The whole system is an *nthimac* because the control can move freely (see Fig. 28). In this case, the remote is a *free component* of the system that also includes the stereo. Moving the remote control is *not tangled* with moving the whole system or the stereo.

3.  *The left half is your part of the cake*. (see Fig. 29) The cake is an *othimac* as in (1) above. We assume here that the left side has already separated from the original cake.

In the TM representation, thimacs signify higher logical types as in UML classes. Individuals (usually called 'objects') are represented in TM as events. Additionally, the meaning of 'part' is indicated diagrammatically by a box inside a larger box in a similar manner to the diagrammatic representation of subsets in set theory.

Note that cutting the cake produces two sides. This fact is implicit in the statement, *The left half is your part of the cake*. The events model illustrates there are three events. Fig. 30 shows two simultaneous events of the parts' appearance. Fig 31 shows the behavior of cutting the cake.

## VI. THE ISSUES RELATED TO THE WHOLE–PART RELATIONSHIP

### A. Priority of the Whole vs. Part

Schaffer [17] considers the issue of priority of the whole versus the part using a circle and a pair of its semicircles. Several questions are raised in this context. Are the semicircles dependent abstractions from their whole, or is the circle a derivative construction from its parts? Schaffer [17] continues, instead of the circle and its semicircles, consider the entire cosmos and the myriad particles. Which, if either, is ultimately prior—the one ultimate whole or its many ultimate parts? The debate about such an issue has long occupied the philosophical center stage as the most central of all philosophic problems. Discussing these questions regarding diagrammatic modeling would illustrate the whole–part problem and clarify the modeling notation.

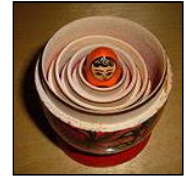

Fig. 26 Matryoshka dolls' movement is only through the whole. Image is from Wikipedia.

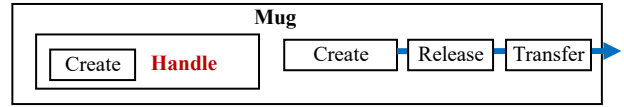

Fig. 27 The part moves wherever the whole moves

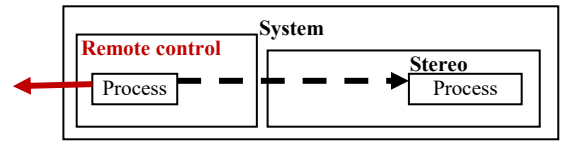

Fig. 28 The part moves freely. Note that create is not shown under the assumption that the box indicates existence in the model. Release and transfer are not shown under the assumption that the direction of the arrow indicates the direction of the movement.

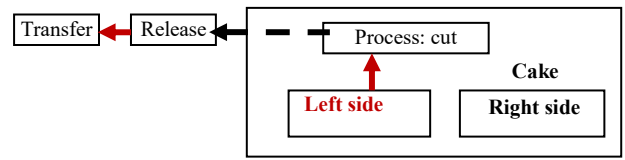

Fig. 29 The part is processed at the whole level to separate it from the whole. Note that the process triggers release and transfer because the left half is already created inside the cake. It is analogous to a bus delivering its passenger.

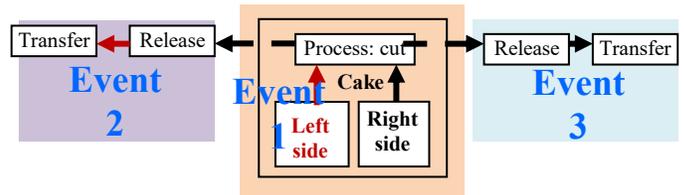

Fig. 30 The events in producing *The left half is your part of the cake*.

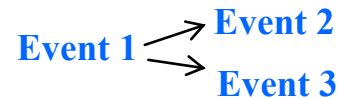

Fig. 31 The behavioral model implicit in *The left half is your part of the cake*.

It is interesting to analyze the classical philosophical issue using the tuple/attribute example in Section 2. In general, the question is whether subthimacs (parts) are dependent abstractions from their thimac (whole), or if the thimac (whole) is a derivative construction from its subthimacs (parts). Usually the issue is discussed regarding the link of the entities that are more fundamental to those that are less.

In the TM approach, the appearances of thimacs in the conceptual model are to be accounted for by reference to the priority (before/after) of appearances of each other. The new idea in the TM approach is the explicit presence of behavior through the machine in the thimac's definition. In this context, the whole–part relationship is based on the thimacs' dependence regarding both entity-ness (thingness) and behavior (machines). The interest here is whether the presence of machines in the dual definition of thimacs affects such an issue. We have to distinguish between *create* as a machine action and the *manifestation* of a thing in a thimac. A machine can be a subdiagram of Fig. 4 (Section 2), that is, not include all five actions, or it can be a complex of these machines.

In TM, the action *create* is in the machine side of the thimac and it corresponds to *manifest* on the thing side (see Fig. 5 in Section 2). Accordingly, when thinking about a thimac (whole) and its subthimacs (parts), we have to separate the *manifestation* of the thing from the *creation* of the machine in any thimac or subthimac. *If* we postulate the *existence* of the part prior to the whole, we have to make the part machines construct the whole machine in the absence of the holistic process that handles all parts (see Fig. 32). We need this holistic process to create a whole, which is missing because of our postulate. Accordingly, we dismiss such a postulate and we adopt the position that the whole exists before the parts (Fig. 33). "Exists" here does not mean ontological existence; rather, it is recognition of what is in the model (see the paragraph about Carnap's existence of multiple types of logical frameworks). According to this position, the whole and the parts are ontologically "there." The whole is "sensed" (e.g., through natural language) and a search is conducted to identify its parts.

### B. Material Objects, Change and Events

It is typical to assume that material objects endure in time without being extended in time, which is in contrast to such phenomena as events, states and processes (as defined in a non-TM context), and are spatially and temporally extended. The argument for such a claim is that a thing is temporally extended if it has parts that the time of their existence distinguishes them. In the TM view, the thimac is continually acting as a machine (e.g., Russell's process of Cleopatra's needle). Such actions involve changes (processes) over time. Thus, every thimac involving time is an event including material objects. Such an issue is usually discussed regarding the difference between *continuants* and events (*occurrences*). Continuants persist through change, whereas events do not. This claim is challenged in some of the philosophical literature on the grounds that no concrete object retains its identity through time; hence, ontology contains occurrences only. In TM, the change is continuous through the thimac's machine without ceasing to exist. Existence is considered as a process in time. As soon as the thimac (of the object, e.g., Cleopatra's needle) appears (exists), time causes processing of existence (creation), thus extending (see Fig. 34) creation (existence).

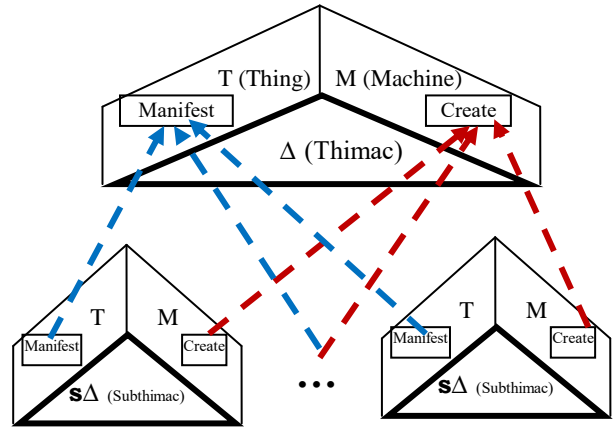

Fig. 32 If the parts' machines construct the whole machine, then this means a holistic beehaior by the parts in the absence of their holistic control

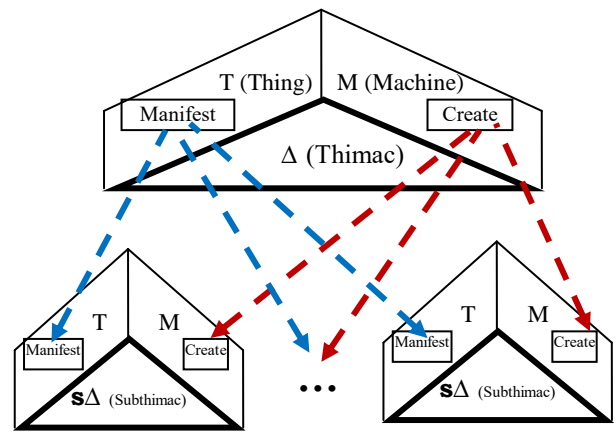

Fig. 33 The thimac manifests its subthimacs and the thimac machine constructs the subthimacs' submachine

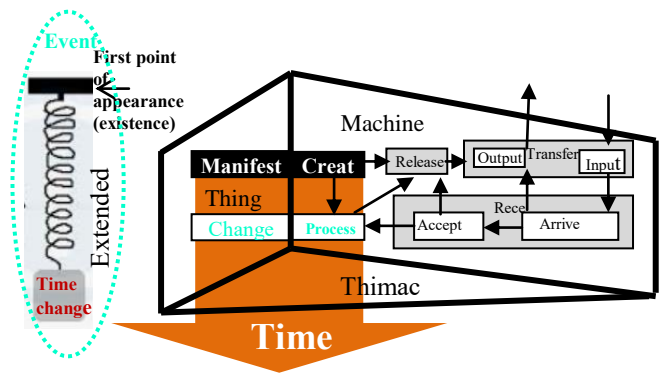

Fig. 34 The thing (instance) extended in time as an event.

**Example:** Simons [18] discusses continuants that come into being, change, and pass. A cat consists of a body and a tail. However, the cat with a tail and the cat without a tail are the same cat without having all parts in common.

Fig. 35 shows the TM representation of this cat example in terms of occurrences (events). Fig. 35 (a) expresses that *there is a cat that exists at a certain period of time and has the indicated parts: body and tail*. Fig. 35 (b) expresses that *the cat loses its tail in an accident*. Fig. 35 (c) expresses that *the cat now does not have a tail*. Fig. 36 shows the sequence of these events. The cat's identity is preserved by this *chronology of events*.

## VII. Conclusion

This paper contributes to establishing a broad ontological foundation for conceptual modeling, specifically, for the 'whole part' (*assembly*) relationship. We addressed the semantics of the whole–part notion by specifying a diagrammatical based on thinging machines representation. To demonstrate the viability of the TM approach, examples of UML modeling were remodeled. The results show a clearer and richer representation of the whole–part relationship and elate notions.

Of course, examining such an extensive relationship is a continuing process that needs further research. Further research will involve experimenting with the TM approach in more examples and using it in various application areas.

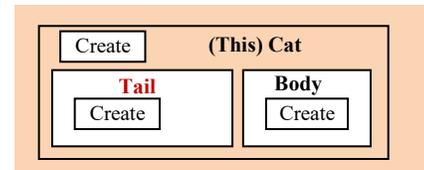

(a) The event: This cat exists in a certain time and has two parts

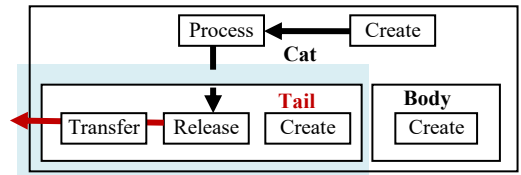

(b) The event: The cat, through an accident, loses its tail

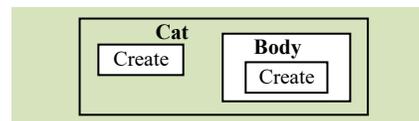

(c) The event: The cat is now without a tail

Fig. 35 The cat representation through events (occurrences)

$$a \longrightarrow b \longrightarrow c$$

Fig. 36 The dynamic model of events (occurrences)